\documentclass[aps]{revtex4}
\usepackage{graphicx}
\usepackage{amsmath}
\usepackage{amssymb}
\begin{document}

\title{Slowly Rotating Dilaton Black hole In Anti-de Sitter Spacetime}
\author{Tanwi Ghosh\footnote{E-mail: tanwi.ghosh@yahoo.co.in} and 
Soumitra SenGupta\footnote{E-mail: tpssg@iacs.res.in}}
\affiliation{Department of Theoretical Physics, Indian Association for 
the
Cultivation of Science,\\
Jadavpur, Calcutta - 700 032, India}
\begin{abstract}
Rotating dilaton black hole solution for asymptotically 
anti-de Sitter spacetime are obtained in the small angular momentum limit 
with an appropriate combination of three Liouville-type dilaton potentials.
The angular momentum, magnetic dipole moment and the gyromagnetic ratio of such a black hole are determined
for arbitrary values of the dilaton-electromagnetic coupling parameter. 
.
\end{abstract}
\maketitle

{\bf I. INTRODUCTION:}\\

Scalar coupled black hole solutions with different asymptotic spacetime 
structure 
is a subject of interest for a long time. There has been a renewed 
interest in such studies ever since new black hole 
solutions have been found in the context of string theory.
The low energy effective action of string theory contains two massless 
scalars namely dilaton and axion.
The dilaton field couples in a non-trivial way to other fields such as 
gauge fields and results into interesting solutions for the 
background spacetime \cite{Garfinkle}. It is found that the dilaton changes the causal 
structure of spacetime and leads to 
curvature singularities at finite radii. Subsequently various other properties have 
also been explored for such scalar coupled black holes
\cite{Gibbons,Brill,Gregory,Koikawa,Boulware,Rakhmanov,Harms,Holzhey}.
Most of these scalar coupled black hole solutions however were asymptotically flat. 
Gao and Zhang \cite{Gao} obtained the asymptotically non-flat charged dilaton black hole 
solutions in anti-de Sitter and de Sitter metric in Schwarzschild coordinate using 
the combination of three Liouville type dilaton potentials. Such potential may arise
from the compactification of a higher dimensional supergravity model \cite{Giddings, Radu}
which originates from the low energy limit of a background string theory.

At this time,
there was also a growing interest to study the rotating 
black hole solutions in presence of dilaton coupled electromagnetic
field in the background.    
An exact solution for a rotating black hole with a special 
dilaton coupling was derived using the inverse scattering method 
\cite{Belinsky}. In an alternative approach Horne and Horowitz 
developed a simpler perturbative technique to find such rotating black hole solution where  a 
small angular momentum is introduced as a perturbation in a non-rotating system  
in presence of a general dilaton-electromagnetic coupling. They studied the properties of
asymptotically flat charged dilaton black holes in the limit of  
infinitesimally small angular momentum \cite{Horne,Shiraishi}. A 
Similar work
then followed by introducing an infinitesimal small charge to the black hole\cite{Casadio}.
Following the method adopted in \cite{Horne}, asymptotically non-flat 
rotating 
dilaton black holes are obtained in \cite{Tanwi} for small value of the 
angular momentum. In a different context M.H. Dehghani et al
obtained charged rotating dilaton black string solutions 
\cite{Dehghani}. However charged rotating dilaton 
black hole solutions for an arbitrary dilaton-electromagnetic 
coupling parameter in a de sitter or anti-de Sitter spacetime is yet to
be found.\\
The interest for studying such dilaton black holes with non-vanishing 
cosmological constant has several reasons.
Gauged supergravity theories in various dimensions are obtained with 
negative cosmological constant in a supersymmetric theory.
In such scenario anti-de Sitter spacetime constitutes the vacuum state 
and the black hole solution in such a spacetime becomes an important area to 
study\cite{Maldacena,Maldacena1,Witten,Klemm,Gubser,Aharnoy}.
Apart from the general interest as a classical solutions of field 
equations in the framework of general relativity,there are also 
cosmological implications of exploring the features of  anti-de sitter and de sitter spacetime 
\cite{Vollick,Ureana,Matos,Padmanabhan,Shin'chi}.The behaviour of galaxies away from 
their centres suggests the presence of a very large
amount of dark matter with $\frac{1}{r^2}$ distribution of energy. One 
can imagine this dark matter in the form of a scalar field such
as a dilaton. Asymptotically flat behaviour is not consistent with the 
observed non-decaying velocity
profile and therefore the Ads or ds solutions are of considerable 
interest in this context. Moreover the observed rotational motion in galaxies 
further suggests the need of studying the rotating metric in such a spacetime.\\
In the backdrop of the scenarios described so far 
it is therefore worthwhile to study the rotating black hole solutions in 
a spacetime with non-zero cosmological constant in presence of dilaton-electromagnetic coupling.
Exponential potentials for scalar fields have been 
used by many 
workers in this context, which closely resembles to Gao-Zhang's model. 
We here study  various features  of a rotating charged 
dilatonic black hole solution in a space with 
non-zero cosmological constant. In particular we concentrate on the anti-de sitter spacetime as considered in \cite{Gao}.
Introducing a small angular momentum, following Horne and Horowitz, 
we aim here  to obtain a rotating black hole solution in anti-de Sitter 
spacetime by generalizing Gao and Zhang's work. We then determine the
expressions for angular momentum, magnetic dipole moment and the gyromagnetic ratio 
for such a black hole.\\

{\bf II. FIELD EQUATIONS AND SOLUTIONS:}\\

We begin with an action of the form, 

\begin{eqnarray}
S= \int 
d^4x\sqrt{-g}[R-2\partial_{\mu}\varphi\partial^{\mu}\varphi-V(\varphi)-e^{-2\alpha\varphi} 
F^2]
\end{eqnarray}
where R is the scalar curvature, $F^2 = F_{\mu\nu}F^{\mu\nu}$ , 
$F_{\mu\nu}$ being  
the usual Maxwell field tensor, $V(\varphi)$ is the potential for 
the dilaton field
$\varphi$ and $\alpha$ measures the strength of the coupling between the 
electromagnetic field and $\varphi$.
While $\alpha = 0$ corresponds to the usual Einstein-Maxwell-scalar 
theory, $\alpha = 1$ indicates the dilaton-electromagnetic
coupling that appears in the low energy string action in Einstein's frame.
Varying the action w.r.t the metric,the Maxwell field and dilaton field 
respectively we obtain the equations of motion as,
\begin{eqnarray}
R_{ab} &=& 2\partial_{a}\varphi\partial_{b}\varphi 
+\frac{1}{2}g_{ab}V+2e^{-2\alpha\varphi}[F_{ac}F_{b}^c-\frac{1}{2}g_{ab}F_{cd}F^{cd}]
\end{eqnarray}
\begin{eqnarray}
\partial_{a}[\sqrt{-g}]e^{-2\alpha\varphi}F^{ab}] &=& 0
\end{eqnarray}
\begin{eqnarray}
\nabla^2\varphi &=& \frac{1}{4}\frac{\partial 
V}{\partial\varphi}-\frac{\alpha}{2}e^{-2\alpha\varphi}F_{cd}F^{cd}
\end{eqnarray}
For arbitrary value of $\alpha$ in anti-de Sitter space the form of the 
dilaton potential is chosen as \cite{Gao},
\begin{equation}
V(\varphi)=\frac{2}{3}\lambda\frac{1}{(\alpha^2+1)^2}[\alpha^2(3\alpha^2-1)e^{-2\varphi/\alpha}
+(3-\alpha^2)e^{2\alpha\varphi}+8\alpha^2e^{\alpha\varphi-\varphi/\alpha}]
\end{equation}
with the metric to be of the form \cite{Gao},
\begin{eqnarray}
ds^2= 
-[(1-\frac{r_+}{r})(1-\frac{r_-}{r})^{\frac{(1-\alpha^2)}{(1+\alpha^2)}}-\frac{1}{3}\lambda 
r^2(1-\frac{r_-}{r})
\nonumber^{\frac{2\alpha^2}{(\alpha^2+1)}}]dt^2 \\
\nonumber+[(1-\frac{r_+}{r})(1-\frac{r_-}{r})^{\frac{(1-\alpha^2)}{(1+\alpha^2)}}-\frac{1}{3}\lambda
 r^2(1-\frac{r_-}{r})^{\frac{2\alpha^2}{(\alpha^2+1)}}]^{-1}dr^2\\
+r^2(1-\frac{r_-}{r})^{\frac{2\alpha^2}{(\alpha^2+1)}}d\Omega^2
-2af(r)sin^2{\theta}dt d\phi
\end{eqnarray}
Here $r_+$ and $r_-$ are respectively the event horizon and Cauchy horizon of the black hole.
The parameter $'a'$ measures the angular momentum which we shall choose 
to be small.
In other words it can be viewed as a small axisymmetric perturbation in 
an anti-de sitter background. 
We now determine the form of f(r) for such an anti-de Sitter metric
and look for the possible 
black hole structure in that spacetime.\\
Integrating the Maxwell's equations we obtain the first 
derivative of the scalar potential as, 
\begin{eqnarray}
-A_0'=F_{01}=\frac{Qe^{2\alpha\varphi}}{h}
\end{eqnarray}
where $Q$, the integration constant is related to the physical electric charge as will be shown later 
and $h$ for the anti-de Sitter metric is given by, 
\begin{eqnarray}
h  =  r^2(1-\frac{r_-}{r})^{\frac{2\alpha^2}{(\alpha^2+1)}} 
\end {eqnarray}
It may be noted that the only term in the metric that changes to the
order of the angular momentum parameter $a$ is
$g_{t\phi}$. It is easy to see that the dilaton solution does not change 
to the order of $a$ and
the solution for the $\phi$ component of the vector potential 
in the presence of rotation is given by, 
\begin{eqnarray}
A_{\phi}=-a Qc(r)sin^2{\theta}
\end{eqnarray}
where c(r) satisfies
\begin{eqnarray}
Q(uc'e^{-2\alpha\varphi})'-\frac{2Qce^{-2\alpha\varphi}}{h}= 
-Q(\frac{f}{h})'
\end{eqnarray}
with,
\begin{eqnarray}
u = [(1-\frac{r_+}{r})(1-\frac{r_-}{r})^{\frac{(1-\alpha^2)}{(1+\alpha^2)}}-\frac{1}{3}\lambda 
r^2(1-\frac{r_-}{r})
\nonumber^{\frac{2\alpha^2}{(\alpha^2+1)}}]
\end{eqnarray}

For anti-de sitter spacetime the solution for the dilaton field is \cite{Gao},
\begin{eqnarray}
e^{2\alpha\varphi}= 
e^{2\alpha\varphi_0}(1-\frac{r_-}{r})^{\frac{2\alpha^2}{(\alpha^2+1)}}
\end{eqnarray}
The remaining equation of motion which is needed to be satisfied for small 
$a$ is, 
\begin{eqnarray}
\frac{f''}{f}=\frac{h''}{h}-\frac{4Q^2c'}{fh}
\end{eqnarray}
For small value of $a$, using the ansatz  $c(r)=\frac{1}{r}$ \cite{Tanwi,Caldarelli}, 
the solution of f(r) from equation (12) and (15)turns out to be,
\begin{eqnarray}
f(r)&=&\frac{r^2(1+\alpha^2)^2 
e^{-2\alpha\varphi_0}}{(1-\alpha^2)(1-3\alpha^2)r_-^2}(1-\frac{r_-}{r})^{\frac{2\alpha^2}{(\alpha^2+1)}}\\
&&-e^{-2\alpha\varphi_0}(1-\frac{r_-}{r})^{\frac{(1-\alpha^2)}{(1+\alpha^2)}}   
[1-\frac{r_+}{r}+\frac{(1-\alpha^2)}{(1+\alpha^2)}\frac{r}{r_-}+\frac{r^2(1+\alpha^2)^2}{(1-\nonumber\alpha^2)(1-3\alpha^2)r_-^2}] 
\\ 
&&+\frac{\lambda}{3}e^{-2\alpha\varphi_0}r^2(1-\frac{r_-}{r})^{\frac{2\alpha^2}{(\alpha^2+1)}}
\end{eqnarray}
It is straightforward to verify that in our case for $\lambda=0$, the 
expression for f(r) agrees with the corresponding expressions 
for a slowly rotating dilaton
black hole solutions in a flat spacetime as obtained in 
\cite{Horne,Shiraishi}.
For slowly rotating dilaton black hole,we learn from \cite{Horne}
that the surface gravity and the area of the event horizon do not 
change to the order of the angular momentum parameter $a$. In the linear approximation in rotation parameter 
$a$ we obtain the 
expression for angular momentum for anti-de Sitter case using Komar 
approach \cite{Aliev}. For rotating Ads black hole, angular momentum and 
mass are
defined by 
\begin{eqnarray}
M'=-\frac{1}{8\pi}\oint *d(\delta\widehat{\xi}_{(t)})
\end{eqnarray}
and
\begin{eqnarray}
J'=\frac{1}{16\pi}\oint*d(\delta\widehat{\xi}_{\phi})
\end{eqnarray}
where *denotes the Hodge dual and $\xi$ the Killing one 
form. $(\delta\widehat{\xi})$ represents the
difference between the Killing isometries of the spacetime and its 
background spacetime. Using the above two integrands in the asymptotic limit, 
we obtain,
\begin{eqnarray}
\delta\xi^{t;r}_{(t)}=\frac{r_+}{2r^2}+(\frac{1-\alpha^2}{1+\alpha^2})\frac{r_-}{2r^2}+O(\frac{1}
{r^2})
\end{eqnarray}
and
\begin{eqnarray}
\delta\xi^{t;r}_{(\phi)}= 
e^{-2\alpha\varphi_0}\frac{r_+}{r^2}+e^{-2\alpha\varphi_0}\frac{r_-}
{r^2}\frac{(3-\alpha^2)}{3(\alpha^2+1)} + O(\frac{1}{r^2})
\end{eqnarray}
Integrating (19) and (20) over two sphere in 
$r\rightarrow \infty$ limit we get
\begin{eqnarray}
M' = \frac{r_+}{2}+(\frac{1-\alpha^2}{1+\alpha^2})\frac{r_-}{2}
\end{eqnarray}
and
\begin{eqnarray}
J'= 
\frac{ae^{-2\alpha\varphi_0}}{2}[r_++\frac{(3-\alpha^2)}{3(1+\alpha^2)}r_-]
\end{eqnarray}
For the anti-de Sitter dilatonic black hole
the expression for actual mass can be written as \cite{Aliev}
\begin{eqnarray}
M_{actual} = M'-a\frac{\lambda}{3}J'=M'-a^2\frac{\lambda}{3}\frac{e^{-2\alpha\varphi_0}}{2}[r_++\frac{(3-\alpha^2)}{3(1+\alpha^2)}r_-]= M'
\end{eqnarray}
Here for slowly rotating black hole we have ignored the term of the order of $a^2$.

It is interesting to note that the expression for J is  same as that 
obtained in \cite{Horne}.
Using the expression for $h$ (from equation 8) and expression for $e^{2\alpha\phi}$ ( from equation 11) in 
equation (7) and (9) one obtains the following expressions for the  
radial fields \cite{Caldarelli},
\begin{eqnarray}
E_{r}=\frac{Qe^{2\alpha\varphi_0}}{r^2}
\end{eqnarray}
and
\begin{eqnarray}
B_{r}=\frac{aQsin^2{\theta}}{r^2}
\end{eqnarray}
The Gaussian flux of electric field gives the value of electric charge 
as follows,
\begin{eqnarray}
Q' = \frac{1}{4\pi}\oint*F= Q e^{2\alpha\varphi_0}
\end{eqnarray}
Magnetic dipole moment for this slowly rotating dilaton black hole in 
anti-de Sitter spacetime can
be defined as
\begin{eqnarray}
\mu'=Q'a=g\frac{Q'J'}{2M'}
\end{eqnarray}
Substituting $M'$, $J'$ and $Q'$ from equ.(17), (18), (19)and (22), g can be obtained as
\begin{eqnarray}
g = 2-\frac{4\alpha^2 r_-}{[(3-\alpha^2)r_-+3(1+\alpha^2)r_+]}
\end{eqnarray}   
It may be noted that while using equ.(20) to estimate  $g$ above , we have ignored the term quadratic in 
the rotation parameter $a$. As a result in the linear approximation in $a$, the above expression for $g$ turns out to
be same as  that found in \cite{Horne}.   
Moreover it is easy to check that for $\alpha = 0$ ,we retrieve g=2 and in the limit
$\alpha \rightarrow 1$ the value of $g$ lies between $2$ and $3/2$.
\\

{\bf IV. CONCLUDING REMARKS:}\\
Starting from the  non-rotating charged dilaton black hole solutions 
in anti-de Sitter spacetime, we have obtained the 
solution for the 
rotating charged dilaton black hole by introducing a small angular 
momentum following the perturbative method suggested in \cite{Horne}. Since it 
is well-known, that in the presence of 
one Liouville type dilaton potential, no de Sitter or anti-de Sitter 
dilaton
black hole exists even in absence of rotation therefore considering the proposal in \cite{Gao}  
of three Liouville type dilaton 
potential we show that such potential 
leads to slowly rotating charged dilaton
black holes solutions in an anti-de Sitter  
spacetime.
As expected, our solutions f(r) reduces to \cite{Horne} for 
$\lambda=0$. The forms of 
angular momentum $J$ and gyromagnetic ratio $g$ are obtained. Their values to the linear
order in the angular momentum parameter $a$ 
turn out to be same as that obtained in \cite{Horne}. It is interesting to note that 
the dilaton field modifies
the value of g from 2 through the coupling parameter $\alpha$ which measures the strength of
the dilaton-electromagnetic coupling.\\
As discussed earlier, the presence of such anti-de sitter dilatonic charged rotating black hole is 
inevitably associated with an 
accompanying scalar field with appropriate Lioville-type potential. Their study 
therefore may lead to a better understanding of the 
origin of the dark matter in the universe. This work brings out the corresponding spacetime solution along with
the associated black hole structure in the limit of small angular momentum. 

{\bf ACKNOWLEDGMENT:}\\

TG wishes to thank CSIR (India) for financial support.\\

\end{document}